\documentclass[11pt]{article}
\usepackage{times,amsmath,amsthm,amssymb,graphicx,hyperref,float,xcolor}
\usepackage{fancyhdr}     
\usepackage{setspace}    
\usepackage{booktabs}    
\usepackage{longtable}   
\usepackage{multirow}
\usepackage{textcomp}    
\usepackage{bm}         
\usepackage{pdfpages}    
\newtheorem{nt}{Note}

\newtheorem{xmpl}{Example}

\newtheorem{case}{Case}

\newcommand{\R}{\mathbb{R}}

\newcommand{\ds}{\displaystyle}

\begin{document}
\global\def\refname{{\normalsize \it References:}}
\baselineskip 12.5pt
%
%
%
\title{\LARGE \bf Bilayer one-dimensional Convection–Diffusion–Reaction-Source problem:
Analytical and numerical solution}

\date{}

\author{\hspace*{-10pt}
\begin{minipage}[t]{2.3in} \normalsize \baselineskip 12.5pt
\centerline{GUILLERMO FEDERICO UMBRICHT}
\centerline{Departamento de Matem\'atica, Facultad de Ciencias Empresariales, Universidad Austral}
\centerline{Paraguay 1950, Rosario, Santa Fe, ARGENTINA}
\centerline{Consejo Nacional de Investigaciones Cient{\'i}ficas y T\'ecnicas (CONICET)}
\centerline{Godoy Cruz 2290, CABA, ARGENTINA}
\vspace{0.5cm}
\centerline{DIANA RUBIO}
\centerline{ITECA (UNSAM-CONICET), CEDEMA, ECyT, Universidad
Nacional de General San Mart\'in}
\centerline{25 de mayo y Francia, San Mart\'in, Buenos Aires, ARGENTINA}
\vspace{0.5cm}
\centerline{DOMINGO ALBERTO TARZIA}
\centerline{Departamento de Matem\'atica, Facultad de Ciencias Empresariales, Universidad Austral}
\centerline{Paraguay 1950, Rosario, Santa Fe, ARGENTINA}
\centerline{Consejo Nacional de Investigaciones Cient{\'i}ficas y T\'ecnicas (CONICET)}
\centerline{Godoy Cruz 2290, CABA, ARGENTINA}
\end{minipage}
%
%
\\ \\ \hspace*{-10pt}
\begin{minipage}[b]{6.9in} \normalsize
\baselineskip 12.5pt {\it Abstract:}
This article presents a theoretical analysis of a one-dimensional heat transfer problem in two layers involving diffusion, advection, internal heat generation or loss linearly dependent on temperature in each layer, and heat generation due to external sources. Additionally, the thermal resistance at the interface between the materials is considered. The situation of interest is modeled mathematically, explicit analytical solutions are found using Fourier techniques, and a convergent finite difference scheme is formulated to simulate specific cases. The solution is consistent with previous results. A numerical example is included that shows coherence between the obtained results and the physics of the problem. The conclusions drawn in this work expand the theoretical understanding of two-layer heat transfer and may also contribute to improving the thermal design of multilayer engineering systems.
\\ [4mm] {\it Key--Words:}
Heat transfer, Multilayer, Composite materials, Interfacial thermal resistance.\\
%
\end{minipage}
\vspace{-10pt}}

\maketitle

\thispagestyle{empty} \pagestyle{empty}
%
%

\section*{Nomenclature}

\begin{longtable}{p{11mm} c p{120mm} }
\multicolumn{3}{l}{\textbf{Subscripts and Superscripts}}\\
\\
$0$ & --- & initial value\\
$m \, (1,2) $ & --- & layer number\\
$n$ & --- & eigenvalue number\\
$H$ & --- & homogeneous system\\
$\infty$ & --- & stationary state\\
$i$ & --- & spatial grid position (numerical method) \\
$j$ & --- & time grid position (numerical method) \\

\\
\multicolumn{3}{l}{\textbf{Capital Letters}}\\
\\

$A$ & --- & auxiliary dimensionless parameter\\
$B$ & --- & auxiliary dimensionless parameter\\
$\bar{A}$ & --- & auxiliary temporal function\\
$\bar{B}$ & --- & auxiliary temporal function\\
$Bi$ & --- & Biot number \\
$Bi^*$ & --- & auxiliary dimensionless parameter \\
$\bar{Bi}$ & --- & auxiliary dimensionless parameter\\
$C$ & --- & the specific heat at constant pressure $\textbf{[J(kg{$^{\circ}$}C)$^{-1}$]}$\\
$D$ & --- & differential operator $\textbf{[{$^{\circ}$}C\,s$^{-1}$]}$\\
$\bar{D}$ & --- & dimensionless differential operator\\
$L$ & --- & body length $\textbf{[m]}$\\
$K$ & --- & auxiliary dimensionless parameter\\
$Pe$ & --- & Péclet number \\
$P$     & --- & auxiliary function (numerical method) $\textbf{[{$^{\circ}$}C]}$\\
$\mathcal{P}$     & --- & partition (numerical method)\\
$R$ & --- & thermal resistance $\textbf{[m]}$\\
$\bar{R}$ & --- & dimensionless thermal resistance\\
$S$    & --- & auxiliary dimensionless heat source\\
$T$    & --- & temperature field relative to ambient $\textbf{[{$^{\circ}$}C]}$\\
$T_r$ & --- & reference temperature $\textbf{[{$^{\circ}$}C]}$\\
$Z$     & --- & auxiliary parameter (numerical method)\\

\\
\multicolumn{3}{l}{\textbf{Lowercase Letters} }\\
\\
$f$ & --- & dimensionless auxiliary spatial function\\
$g$ & --- & dimensionless auxiliary temporal function\\
$l$ & --- & interface location $\textbf{[m]}$\\
$\bar{l}$ & --- & dimensionless interface location\\
$q$ & --- & auxiliary function\\
$r$ & --- & auxiliary function\\
$s$  & --- & heat source $\textbf{[{$^{\circ}$}C\,s$^{-1}$]}$\\
$\bar{s}$ & --- & dimensionless auxiliary heat source\\
$\widehat{s}$ & --- & dimensionless heat source\\
$h$     & --- & convection heat transfer coefficient $\textbf{[Wm{$^{-2}$}({$^{\circ}$}C)$^{-1}$]}$\\
$t$     & --- & temporary variable $\textbf{[s]}$\\
$x$     & --- & spatial variable $\textbf{[m]}$\\
$y$     & --- & dimensionless spatial variable\\

\\
\multicolumn{3}{l}{\textbf{Greek Letters}}\\
\\

$\alpha$ & --- & thermal diffusivity coefficient $\textbf{[m{$^{2}$}s$^{-1}$]}$\\
$\bar{\alpha}$ & --- & dimensionless thermal diffusivity coefficient\\
$\beta$ & --- & fluid velocity $\textbf{[m\,s$^{-1}$]}$\\
$\nu$ & --- & generation/consumption coefficient $\textbf{[s$^{-1}$]}$\\
$\bar{\nu}$ & --- & dimensionless generation/consumption coefficient\\
$\kappa$ & --- & thermal conductivity coefficient $\textbf{[W(m{$^{\circ}$}C)$^{-1}$]}$\\
$\bar{\kappa}$ & --- & dimensionless thermal conductivity coefficient\\
$\rho$ & --- & density $\textbf{[kg m$^{-3}$]}$\\
$\tau$    & --- & dimensionless temporary variable\\
$\theta$ & --- & dimensionless temperature\\
$\theta$ & --- & dimensionless auxiliary temperature function\\
$\chi$ & --- & auxiliary dimensionless parameter\\
$\Delta t$     & --- & time discretization step (numerical method) $\textbf{[s]}$\\
$\Delta x$     & --- & spatial discretization step (numerical method) $\textbf{[m]}$\\
$\gamma$ & --- & auxiliary dimensionless parameter\\
$\sigma$ & --- & auxiliary dimensionless parameter\\
$\epsilon$     & --- & auxiliary parameter (numerical method)\\
$\varphi$    & --- & auxiliary dimensionless parameter\\
$\psi$     & --- & auxiliary dimensionless parameter\\
$\mu$     & --- & auxiliary dimensionless parameter\\
$\phi$ & --- & auxiliary dimensionless parameter\\
$\xi$ & --- & auxiliary dimensionless parameter\\
$\eta$ & --- & auxiliary dimensionless parameter\\
$\delta$ & --- & auxiliary dimensionless parameter\\
$\lambda$     & --- & dimensionless temporal eigenvalue\\
$\omega$     & --- & dimensionless spatial eigenvalue\\
$\Omega$     & --- & auxiliary parameter (numerical method)\\
$\Lambda$ & --- & auxiliary parameter (numerical method) $\textbf{[W(m{$^{\circ}$}C)$^{-1}$]}$\\
$\upsilon$     & --- & auxiliary parameter (numerical method)\\
$\Pi$     & --- & auxiliary parameter (numerical method) $\textbf{[m$^{-1}$]}$\\
$\zeta$     & --- & auxiliary parameter (numerical method)\\
\end{longtable}

\section{Introduction}
\label{s:1} \vspace{-4pt}

The mathematical modeling of heat and mass transfer problems in multilayer materials has been extensively studied recently \cite{Yuan22,Carson22,Zhou21,Yavaraj23} due to the numerous applications \cite{Hickson09} in various fields of science, engineering, and industry. The variety of applications is evident from the numerous articles found  in the literature across different disciplines. For instance, wool cleaning techniques \cite{Caunce08}, pollution in porous media \cite{Liu98,Liu08}, skin permeability \cite{Mitragotri11}, drug release analysis in stents \cite{McGinty11}, greenhouse gas emissions \cite{Liu09}, moisture in composite tissues \cite{Pasupuleti11}, thermal conduction in composite materials \cite{Monte00}, brain tumor growth \cite{Mantzavinos16}, heat conduction through the skin \cite{Becker12}, analysis of lithium-ion cells \cite{Bandhauer11}, microelectronics \cite{Choobineh15}, among others.

A relevant and up-to-date state of the art in multilayer material transfer and the mathematical techniques used can be found in \cite{Monte00,Monte02,Jain21}. These problems have been analytically addressed by different methods, among them a recursive images method \cite{Dias15}, the method of separation of variables \cite{Zhou21,Hickson09,Monte00,Monte02,Ma04,Rubio21}, the solution using integral functions such as Laplace and Fourier transform \cite{Goldner92,Dudko04,Simoes05,Rodrigo16}. Numerical techniques have also been used, such as the method of fundamental solutions \cite{Johansson09}, finite differences, and finite elements \cite{Yuan22,Hickson09,Rubio21}.

Although the bibliography is extensive, it lacks generality because the models presented are not complete. Most of the articles cited consider only diffusion, neglecting dissipative terms and sources of the complete parabolic equation. Additionally, most of them do not account for the resistance offered by the interface. For instance, in \cite{Jain21, the authors provide a thorough analysis of heat transfer in multilayer materials but omit the effects of external sources and thermal contact resistance.} Other articles consider heat transfer problems in multilayer materials but only take into account the steady state \cite{Umbricht22,Rubio22,Umbricht22b,Umbricht21,Umbricht20}.

It is important to analyze the influence of external heat generation sources, dissipative terms, and thermal contact resistance, as these are key physical processes in multilayer mass and heat transfer problems. These processes include diffusion, advection, internal heat generation or consumption, and heat generation due to external sources. The rate of internal heat generation or consumption is often considered proportional to the local temperature. Some processes modeled this way include a chemical reaction with first-order kinetics \cite{Shah16,Esho18}, the perfusion term in the Pennes bioheat transfer equation \cite{Pennes48}, and the fin equation used for the analysis of a multilayer segmented fin \cite{Becker13}. The advective term plays an important role in several transfer processes, such as in a flow battery \cite{Skyllas11}. Furthermore, the source term is useful for modeling different processes where external heat is delivered to the system \cite{Kim20}.

This work proposes a mathematical study of transient heat transfer in a bilayer body governed by a Convection-Diffusion-Reaction-Source (CDRS) equation. Diffusion, advection, internal heat generation or loss, heat generation from external sources, and thermal resistance by contact offered by the interface are considered. An analytical expression for the solution to the studied problem is obtained, which is consistent with previous results. Additionally, the proposed numerical approach aims to simulate solutions for specific case studies using finite difference methods.
\section{Mathematical Modeling}

The problem of interest focuses on the transport or transient transfer of heat in a one-dimensional bilayer body. The material in each layer is assumed to be homogeneous and isotropic. Additionally, heat gain or loss within each layer at a rate proportional to the local temperature and advection driven by a one-dimensional fluid flow are taken into account. Furthermore, heat generation from external sources is included. The phenomena of thermal runaway and heat transfer by radiation are ignored.

\begin{figure}[h!]
\begin{center}
\includegraphics[width=0.50\textwidth]{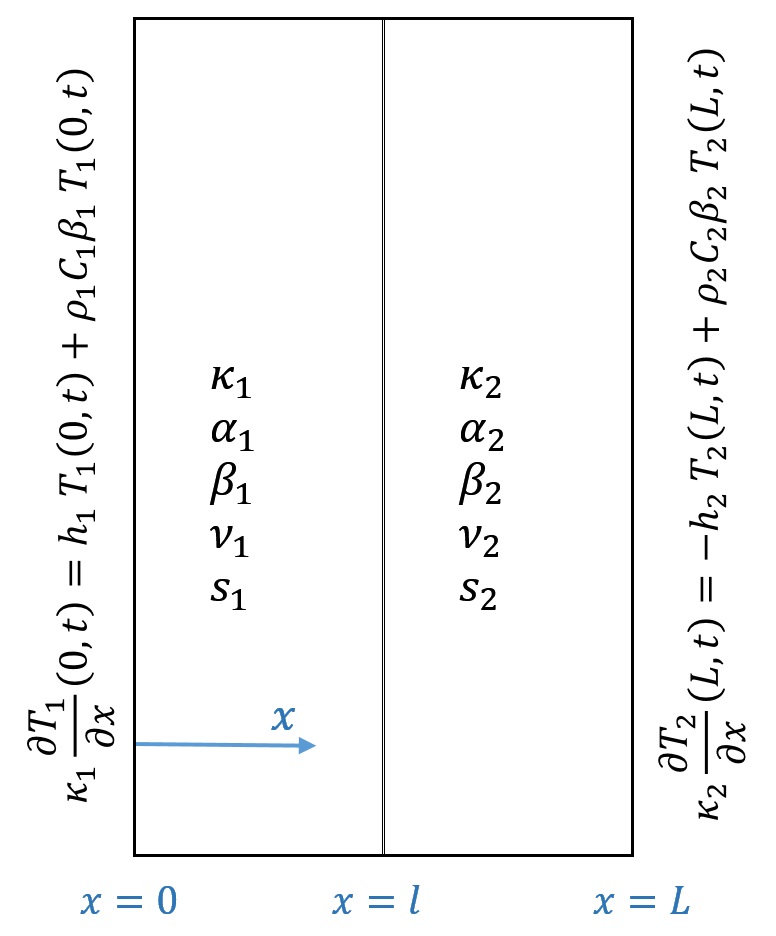}
\caption{General scheme of the problem of interest.}
\label{Esq_Gral}
\end{center}
\end{figure}

The length of the bilayer body is denoted as $L$. The interface, which represents the junction between the two materials or layers, is positioned at $l$, where $L > l$. In Fig. \ref{Esq_Gral}, the reference diagram is shown, where the arrow indicates the direction of heat flow.

A transient energy conservation equation, representing a balance between diffusion, advection, internal heat gain or loss, and heat generation from external sources in a one-dimensional bilayer body, can be written as:
\begin{equation}
\label{Ec_Parab}
\begin{cases}
\dfrac{\partial{T_1}}{\partial{t}}(x,t)= D_1 T_1(x,t) + s_1(x,t), & \quad (x,t) \in (0,l) \times \R^+,  \\
\\
\dfrac{\partial{T_2}}{\partial{t}}(x,t)=D_2 T_2(x,t) + s_2(x,t), & \quad (x,t) \in (l,L) \times \R^+, 
\end{cases}
\end{equation}
where $D_m$ is a parabolic differential operator that has already been used in other works \cite{Umbricht21b} and is defined by
\begin{equation}
\label{Operator_D}
D_m T_m(x,t):= \alpha_{m} \, \dfrac{\partial^2{T_m}}{\partial{x^2}}(x,t)-\beta_m \, \dfrac{\partial{T_m}}{\partial{x}} (x,t)+ \nu_m \, T_m(x,t), \quad m=1,2. 
\end{equation}

In equations \eqref{Ec_Parab}-\eqref{Operator_D}, often referred to as the CDRS equation, the subscripts refer to the first and second layers of the material, where $x$ and $t$ represent the spatial and temporal variables, respectively. The functions $T_1(x, t)$ and $T_2(x, t)$, which satisfy $T_1(x, t) \in C^2(0,l) \times C^1(0,+\infty)$ and $T_2(x, t) \in C^2(l,L) \times C^1(0,+\infty)$, represent the temperature field, relative to ambient temperature, of the first and second layers, respectively, at position $x$ at time $t$.

The first two terms on the right-hand side of equation \eqref{Operator_D} represent heat transfer due to diffusion and advection, respectively, while the third term accounts for internal heat gain or loss proportional to the local temperature. The coefficient $\alpha_m$ denotes the thermal diffusivity of the material in each layer, $\beta_m$ represents the flow velocity, and $\nu_m$ corresponds to the coefficient that relates the rate of internal heat gain or loss to the local temperature. Finally, the differentiable functions $s_i$ given in \eqref{Ec_Parab} model an external source that delivers heat to the body. All properties are assumed to be temperature-independent. An analogous equation can be used to model the concentration field in a one-dimensional mass transfer problem \cite{Kim20}.

Heat is generated due to the presence of external sources. Additionally, heat is either lost or generated within each layer at a rate proportional to the local temperature. Heat transfer within the body occurs due to diffusion and advection driven by an imposed one-dimensional fluid flow from left to right in each layer. Each layer has distinct thermal properties, flow velocity, and heat generation rate.

General convective boundary conditions are assumed on the left and right boundaries, respectively. These  conditions represent a balance between two factors: convective heat transfer between the body and the surroundings, and the diffusion and advection into and out of the body. Note that while advection transfers energy from the environment to the first layer, it also removes energy from the second layer to the environment.

\begin{equation}
\label{Cond_Borde}
\begin{cases}
\kappa_1 \, \dfrac{\partial{T_1}}{\partial{x}}(x,t)=h_1 \,T_1(x,t) + \rho_1  C_1  \beta_1 \,T_1(x,t) , & \qquad x=0, \,\,\, t \in \R^+, 
\vspace{0.2cm}
\\
\kappa_2 \, \dfrac{\partial{T_2}}{\partial{x}}(x,t)=-h_2 \,T_2(x,t) + \rho_2  C_2  \beta_2 \,T_2(x,t), & \qquad x=L, \,\,\, t \in \R^+,
\end{cases} 
\end{equation}
where $\kappa_m$, $h_m$, $\rho_m$ and $C_m$ denote: the thermal conductivity coefficient, the convection heat transfer coefficient, the density and the specific heat at constant pressure for each layer, respectively.

Additionally, temperature discontinuity at the interface due to thermal contact resistance at the material junction is considered. Typically, this discontinuity is modeled such that the temperature difference between the layers is proportional to the heat flux at the interface \cite{Zhou21b}; this type of thermal jump is accounted for in the first equation of \eqref{Cond_Interf}. Regarding the heat flux at the interface, the principle of energy conservation is applied, leading to the thermal flux equality stated in the second equation of \eqref{Cond_Interf}.
\begin{equation}
\label{Cond_Interf}
\begin{cases}
T_2(x,t)=T_1(x,t)+ R \, \dfrac{\partial{T_1}}{\partial{x}} (x,t), &  x=l, \, t \in \R^+,  \\
\kappa_2  \dfrac{\partial{T_2}}{\partial{x}}(x,t)- \rho_2  C_2  \beta_2 T_2(x,t)=\kappa_1 \dfrac{\partial{T_1}}{\partial{x}}(x,t)- \rho_1  C_1  \beta_1 T_1(x,t), & x=l, \, t \in \R^+, 
\end{cases} 
\end{equation}
where $R$ denotes the value of the thermal contact resistance at the interface between the materials. Finally, an initial spatial distribution of temperature is assumed for each layer. That is to say,

\begin{equation}
\label{Cond_Inicial}
\begin{cases}
T_1(x,t)=T_{1,0} (x),  & \qquad  x \in \left[0,l\right], \,\,\, t=0,  \\
T_2(x,t)=T_{2,0} (x), & \qquad x \in \left[l,L\right], \,\,\, t=0. 
\end{cases} 
\end{equation} 

\begin{nt} 
The problem defined by equations \eqref{Ec_Parab}-\eqref{Cond_Inicial} is studied at a macroscopic scale, as results obtained may not be valid at other scales. This is primarily because the thermophysical properties of the interface between materials and their effects can change significantly at different scales. For instance, at the nanometric scale, the one-dimensional heat transfer problem between two layers cannot be effectively solved using the techniques presented in this work. At this scale, it is necessary to use other methods such as non-equilibrium molecular dynamics simulations or non-equilibrium Green's function calculations based on interatomic potentials. This nanometric-scale problem is relevant in the study of interface nanodevices and has been recently addressed by various authors \cite{Yang19,Li23,Yang24,Yang24b} for different types of materials, including graphene-silver, graphene-gold, graphene-silicon, and graphene-copper.
\end{nt}

In the next section, we obtain an explicit analytical solution to the problem described by the equations \eqref{Ec_Parab}-\eqref{Cond_Inicial}.

\section{Analytical Solution} \label{Solucion_analitica}

The transient heat transfer problem to be solved is defined by equations \eqref{Ec_Parab}-\eqref{Cond_Inicial}. To simplify the problem, the equations are made dimensionless by introducing the following parameters for $m=1,2$
\begin{equation}
\label{Aux0}
\begin{cases} 
y=\dfrac{x}{L}, \quad \bar{l}=\dfrac{l}{L}, \quad \bar{R}=\dfrac{R}{L}, \quad \tau=\dfrac{\alpha_2}{L^2} \,t, \quad \theta_m=\dfrac{T_m}{T_{r}}, \quad \bar{\alpha}=\dfrac{\alpha_1}{\alpha_2}, \quad {Pe}_m=\dfrac{L}{\alpha_2} \, \beta_m,  \quad 
\vspace{0.2cm}\\ \bar{\nu}_m=\dfrac{L^2}{\alpha_2} \,\nu_m, \quad \bar{s}_m= \dfrac{L^2}{T_{r} \, \alpha_2} \, s_m, \quad \bar{\kappa}=\dfrac{\kappa_1}{\kappa_2}, \quad {Bi}_m=\dfrac{L}{\kappa_2} \, h_m, 
\end{cases}
\end{equation}
where $Pe_m$, $Bi_m$ denote the dimensionless Péclet and Biot numbers, respectively, and the parameter $T_r$ represents a reference temperature. This change of variables is applied to equations \eqref{Ec_Parab}-\eqref{Cond_Inicial}, resulting in the following dimensionless system:
\begin{equation}
\label{SisT_adimen}
\begin{cases} 
\dfrac{\partial{\theta_1}}{\partial{\tau}}(y,\tau)= \bar{D}_1 \theta_1 (y,\tau) + \bar{s}_1(y,\tau), \, & (y,\tau) \in (0,\bar{l}) \times \R^+, \vspace{0.1cm}\\
\dfrac{\partial{\theta_2}}{\partial{\tau}}(y,\tau)= \bar{D}_2 \theta_2 (y,\tau) + \bar{s}_2(y,\tau), \, & (y,\tau) \in (\bar{l},1) \times \R^+, \vspace{0.1cm}\\
\dfrac{\partial{\theta_1}}{\partial{y}}(y,\tau)=  {{Bi}_1}^* \, \theta_1(y,\tau) , \, & y=0 , \, \tau \in \R^+, \vspace{0.1cm}\\
\dfrac{\partial{\theta_2}}{\partial{y}}(y,\tau)=  {{Bi}_2}^* \, \theta_2(y,\tau) , \, & y=1 , \, \tau \in \R^+, \\
\theta_2(y,\tau) = \theta_1(y,\tau) + \bar{R} \, \dfrac{\partial{\theta_1}}{\partial{y}}(y,\tau) , \, & y=\bar{l} , \, \tau \in \R^+, \\
\dfrac{\partial{\theta_2}}{\partial{y}}(y,\tau)= \gamma \, \theta_1(y,\tau) + \sigma \, \dfrac{\partial{\theta_1}}{\partial{y}}(y,\tau), \, & y=\bar{l} , \, \tau \in \R^+, \\
\theta_1(y,\tau)=\theta_{1,0} (y),\, & y \in \left[0,\bar{l}\right], \, \tau=0,\\
\theta_2(y,\tau)=\theta_{2,0} (y), \, & y \in \left[\bar{l},1\right], \, \tau=0,
\end{cases}
\end{equation}
where
\begin{equation}
\label{Oper_adimen}
\begin{cases} 
\bar{D}_1 \theta_1 (y,\tau)=\bar{\alpha} \, \dfrac{\partial^2{\theta_1}}{\partial{y^2}}(y,\tau) - Pe_1 \,\dfrac{\partial{\theta_1}}{\partial{y}}(y,\tau) + \bar{\nu}_1 \, \theta_1(y,\tau),  \\
\bar{D}_2 \theta_2 (y,\tau)= \dfrac{\partial^2{\theta_2}}{\partial{y^2}}(y,\tau) - Pe_2 \,\dfrac{\partial{\theta_2}}{\partial{y}}(y,\tau) + \bar{\nu}_2 \, \theta_2(y,\tau)
\end{cases}
\end{equation}
and 
\begin{equation}
\label{Par_nuevos}
{{Bi}_1}^*=\dfrac{Pe_1}{\bar{\alpha}}+\dfrac{{Bi}_1}{\bar{\kappa}},  \quad
{{Bi}_2}^*= Pe_2 - {Bi}_2,\quad 
\gamma= Pe_2- Pe_1 \, \dfrac{\bar{\kappa}}{\bar{\alpha}},\quad
\sigma=\bar{\kappa} + \bar{R}\,Pe_2.
\end{equation}

Then, the advective term is removed from equation \eqref{Oper_adimen} by applying a substitution that can be interpreted as a change in the coordinate system. This transformation effectively shifts the system into a reference frame moving with the fluid velocity. Similar coordinate system changes have been used in the literature to address various situations. For instance, see \cite{Basha93,Bharati17,Das17,Sanskrityayn17}. The proposed substitution in this case is:

\begin{equation}
\label{Def_w}
\begin{cases} 
\theta_1(y,\tau)=\exp\left(\chi_1\,y\right) \, \Theta_1(y,\tau),  \qquad  & (y,\tau) \in [0,\bar{l}] \times \R^+, \vspace{0.2cm}\\
\theta_2(y,\tau)=\exp\left(\chi_2\,y\right) \, \Theta_2(y,\tau),  \qquad  & (y,\tau) \in [\bar{l},1] \times \R^+,
\end{cases}
\end{equation}
where
\begin{equation}
\label{chis}
\chi_1= \dfrac{Pe_1}{2 \,\bar{\alpha}}, \qquad \chi_2= \dfrac{Pe_2}{2}.
\end{equation}
The change of variables \eqref{Def_w}-\eqref{chis} is applied to equations \eqref{SisT_adimen}-\eqref{Par_nuevos} leading to the following system
\begin{equation}
\label{Sist_w1} 
\begin{cases} 
\dfrac{\partial{\Theta_1}}{\partial{\tau}}(y,\tau)=\bar{\alpha} \, \dfrac{\partial^2{\Theta_1}}{\partial{y^2}}(y,\tau) + \psi_1 \, \Theta_1(y,\tau) + \widehat{s}_1(y,\tau), \, & (y,\tau) \in (0,\bar{l}) \times \R^+, \\
\dfrac{\partial{\Theta_2}}{\partial{t}}(y,\tau)= \dfrac{\partial^2{\Theta_2}}{\partial{y^2}}(y,\tau) + \psi_2 \, \Theta_2(y,\tau) + \widehat{s}_2(y,\tau), \, & (y,\tau) \in (\bar{l},1) \times \R^+, \\
\dfrac{\partial{\Theta_1}}{\partial{y}}(y,\tau)=  \bar{{Bi}}_1 \, \Theta_1(y,\tau) , \, & y=0 , \, \tau \in \R^+, \vspace{0.1cm}\\
\dfrac{\partial{\Theta_2}}{\partial{y}}(y,\tau)=  \bar{{Bi}}_2 \, \Theta_2(y,\tau) , \, & y=1 , \, \tau \in \R^+, \\
\Theta_2(y,\tau) = \phi \, \Theta_1(y,\tau) + \mu \, \dfrac{\partial{\Theta_1}}{\partial{y}}(y,\tau) , \, & y=\bar{l} , \, \tau \in \R^+, \\
\dfrac{\partial{\Theta_2}}{\partial{y}}(y,\tau)= \eta \, \Theta_1(y,\tau) + \varphi \, \dfrac{\partial{\Theta_1}}{\partial{y}}(y,\tau), \, & y=\bar{l} , \, \tau \in \R^+, \\
\Theta_1(y,\tau)=\Theta_{1,0} (y),\, & y \in \left[0,\bar{l}\right], \, \tau=0,\\
\Theta_2(y,\tau)=\Theta_{2,0} (y), \, & y \in \left[\bar{l},1\right], \, \tau=0,
\end{cases}
\end{equation}
where
\begin{equation}
\label{Aux1}
\begin{cases} 
\psi_1= \bar{\nu}_1 - \bar{\alpha} \, \chi_1^2, \quad \psi_2= \bar{\nu}_2 - \chi_2^2,  \quad \widehat{s}_1(y,\tau)= \bar{s}_1(y,\tau) \, \exp\left(-\chi_1 \, y \right),\\ 
\widehat{s}_2(y,\tau)= \bar{s}_2(y,\tau) \, \exp\left(-\chi_2 \, y\right), \quad
\bar{{Bi}}_1 = {{Bi}_1}^*-\chi_1, \quad \bar{{Bi}}_2 = {{Bi}_2}^*-\chi_2, \\ \phi= \xi \,\delta , \quad \mu= \xi\,\bar{R} ,\quad \eta=\xi \, \left(\gamma+ \sigma \,\chi_1 - \delta\,\chi_2   \right), \quad 
\varphi=\xi \, \left(\sigma - \bar{R} \,\chi_2 \right), \\ \xi= \exp\left(\, \bar{l} \, (\chi_1-\chi_2)\right), \quad  \delta=1+ \bar{R} \, \chi_1,  \quad \Theta_{1,0} (y)=\theta_{1,0} (y) \, \exp\left(-\chi_1 \, y \right),  \\
\Theta_{2,0} (y)=\theta_{2,0} (y) \, \exp\left(-\chi_2 \, y \right).
\end{cases}
\end{equation}
The homogeneous system associated with \eqref{Sist_w1}-\eqref{Aux1} is considered, i.e., the source terms \(\widehat{s}_1(y,\tau)\) and \(\widehat{s}_2(y,\tau)\) are  excluded. This homogeneous system is addressed using the method of separation of variables. It is assumed that there exist functions $f_{1,n} \in C^2 (0,\bar{l})$, $f_{2,n} \in C^2 (\bar{l},1)$ and $g_n \in C^1 (0,+\infty)$ such that 
\begin{equation}
\label{SepVar1}
\begin{cases} 
\Theta_1^H(y,\tau)= \ds \sum_{n=1}^{\infty} {f_{1,n} (y) \, g_n(\tau)}, \qquad & (y,\tau) \in (0,\bar{l}) \times \R^+, \\ 
\Theta_2^H(y,\tau)= \ds \sum_{n=1}^{\infty} {f_{2,n} (y) \, g_n(\tau)},  \qquad & (y,\tau) \in (\bar{l},1) \times \R^+.
\end{cases}
\end{equation}
By substituting \eqref{SepVar1} into the homogeneous system associated with \eqref{Sist_w1}-\eqref{Aux1}, it can be shown that $g_n (\tau) = K_n \, \exp(-\lambda_n^2 \, \tau)$, where $\lambda_n$ are the eigenvalues and $K_n$ is a sequence associated with the initial temperature value. In addition, the functions $f_{m,n}$ for $m=1,2$ satisfy 
\begin{equation}
\label{fs}
\begin{cases} 
\bar{\alpha} \, f''_{1,n}(y)+\psi_1 \,f_{1,n}(y)= - \lambda_n^2 \,f_{1,n}(y), \qquad & y \in (0,\bar{l}), \\ 
f''_{2,n}(y)+\psi_2 \,f_{2,n}(y)= - \lambda_n^2 \,f_{2,n}(y), \qquad & y \in (\bar{l},1), \\
f'_{1,n}(y)=  \bar{{Bi}}_1 \, f_{1,n}(y) , \, & y=0 , \\
f'_{2,n}(y)=  \bar{{Bi}}_2 \, f_{2,n}(y) , \, & y=1 , \\
f_{2,n}(y) = \phi \, f_{1,n}(y) + \mu \, f'_{1,n}(y) , \, & y=\bar{l}, \\
f'_{2,n}(y) = \eta \, f_{1,n}(y) + \varphi \, f'_{1,n}(y) , \, & y=\bar{l}, 
\end{cases}
\end{equation}
yielding
\begin{equation}
\label{fs2}
\begin{cases} 
f_{1,n}(y)= A_{1,n}  \cos (\omega_{1,n} \, y) + B_{1,n}  \sin (\omega_{1,n}\, y), \qquad & y \in [0,\bar{l}], \\  
f_{2,n}(y)= A_{2,n}  \cos (\omega_{2,n} \,y) + B_{2,n}  \sin (\omega_{2,n} \, y), \qquad & y \in [\bar{l},1].
\end{cases}
\end{equation}
Then, the solutions of the homogeneous system associated with \eqref{Sist_w1}-\eqref{Aux1} are written as
\begin{equation}
\label{Sol}
\begin{cases} 
\Theta_1^H(y,\tau)= \ds \sum_{n=1}^{\infty} { K_n \left[A_{1,n}  \cos (\omega_{1,n} \, y) + B_{1,n}  \sin (\omega_{1,n}\, y) \right] \exp(-\lambda_n^2 \, \tau)},  \\ 
\Theta_2^H(y,\tau)= \ds \sum_{n=1}^{\infty} { K_n  \left[A_{2,n}  \cos (\omega_{2,n} \,y) + B_{2,n}  \sin (\omega_{2,n} \, y) \right] \exp(-\lambda_n^2 \, \tau)},
\end{cases}
\end{equation}
where $\omega_{m,n}$ with $m=1,2$ are the spatial eigenvalues, which are given by
\begin{equation}
\label{Omegas}
\begin{cases} 
\omega_{1,n}=\omega_{1,n}(\lambda_n)=\sqrt{\dfrac{\lambda_n^2+\psi_1}{\bar{\alpha}}}=\sqrt{\dfrac{\lambda_n^2+\bar{\nu}_1-\bar{\alpha} \chi_1^2}{\bar{\alpha}}}=\sqrt{\dfrac{\lambda_n^2+\bar{\nu}_1- \frac{Pe_1^2}{4\bar{\alpha}}}{\bar{\alpha}}}
, \vspace{0.1cm} \\ 
\omega_{2,n}=\omega_{2,n}(\lambda_n)=\sqrt{\lambda_n^2+\psi_2}=\sqrt{\lambda_n^2+\bar{\nu}_2- \chi_2^2}    =\sqrt{\lambda_n^2+\bar{\nu}_2- \frac{Pe_2^2}{4}}.
\end{cases}
\end{equation}
Now, \(A_{m,n}\), \(B_{m,n}\) with \(m=1,2\), and \(\lambda_n\) in \eqref{Sol}-\eqref{Omegas} are determined redusing the boundary and interface conditions from \eqref{fs}. Additionally, it is assumed that the associated homogeneous system has a non-trivial solution. Algebraic operations are performed and the following expressions are obtained $A_{1,n}=1$, $B_{1,n}=\frac{\bar{{Bi}}_1}{\omega_{1,n}}$,
$A_{2,n}=A_n$ and $B_{2,n}=B_n$ where
\begin{equation}
\label{An}
\begin{split}
A_n=& \dfrac{\sin(\omega_{1,n} \, \bar{l})}{\cos(\omega_{2,n} \, \bar{l})} \left(\phi \, \dfrac{\bar{{Bi}}_1}{\omega_{1,n}}-\mu \, \omega_{1,n} \right)
  + \dfrac{\cos(\omega_{1,n} \, \bar{l})}{\cos(\omega_{2,n} \, \bar{l})} \left(\phi +\mu \, \bar{{Bi}}_1 \right) - \tan (\omega_{2,n} \, \bar{l}) \, B_n   
\end{split}
\end{equation}
and
\begin{equation}
\label{Bn}
\begin{split}
 B_n = & \sin(\omega_{2,n} \, \bar{l}) \left[\sin(\omega_{1,n} \, \bar{l}) \left(\phi \, \dfrac{\bar{{Bi}}_1}{\omega_{1,n}} - \mu \, \omega_{1,n} \,\right) + \cos(\omega_{1,n} \, \bar{l}) \left(\phi + \mu \, \bar{{Bi}}_1 \, \right) \right] \\
 +& \dfrac{\cos(\omega_{2,n} \, \bar{l})}{\omega_{2,n}}  \left[\sin(\omega_{1,n} \, \bar{l})\left(\eta \,\dfrac{ \bar{{Bi}}_1}{\omega_{1,n}} - \varphi \, \omega_{1,n}\right) + \cos(\omega_{1,n} \, \bar{l}) \left(\eta+\varphi \, \bar{{Bi}}_1\right)\right].
\end{split}
\end{equation}
The eigenvalues $\lambda_n$ which are discussed in more detail in Section \ref{Autovalores}, are the solutions of the transcendental equation given by:
\begin{equation}
\label{Tang}
\tan (\omega_{2,n})=\dfrac{\omega_{2,n} \, B_n- \bar{{Bi}}_2 \, A_n}{\bar{{Bi}}_2 \, B_n+\omega_{2,n}\,  A_n},
\end{equation}
with $A_n$ and $B_n$ given by \eqref{An} and \eqref{Bn} respectively.

The Fourier method is used to solve the non-homogeneous system \eqref{Sist_w1}-\eqref{Aux1}. It is assumed that two countably infinite sets of time functions, denoted by $\bar{A_n}(\tau)$ and $\bar{B_n}(\tau)$ satisfy
\begin{equation}
\label{SolNoHomegeneo2}
\begin{cases} 
\Theta_1(y,\tau)= \ds \sum_{n=1}^{\infty} {\bar{A_n}(\tau) \, f_{1,n} (y) }, \quad & (y,\tau) \in [0,\bar{l}] \times \R^+, \\  
\Theta_2(y,\tau)= \ds \sum_{n=1}^{\infty}  { \bar{B_n}(\tau) \, f_{2,n} (y) }, \quad & (y,\tau) \in [\bar{l},1] \times \R^+.
\end{cases} 
\end{equation}
where $f_{m,n}$ with $m=1,2$ are defined in \eqref{fs2}. The source functions $\widehat{s}_1(y,\tau)$ and $\widehat{s}_2(y,\tau)$ in \eqref{Sist_w1} are developed in a series of eigenfunctions.
\begin{equation}
\label{Fuenteenserie}
\begin{cases} 
\widehat{s}_1(y,\tau)= \ds \sum_{n=1}^{\infty} { S_{1,n}(\tau) \, f_{1,n} (y) }, \quad & (y,\tau) \in [0,\bar{l}] \times \R^+,  \\ 
\widehat{s}_2(y,\tau)= \ds \sum_{n=1}^{\infty}  { S_{2,n}(\tau) \, f_{2,n} (y) }, \quad & (y,\tau) \in [\bar{l},1] \times \R^+,
\end{cases} 
\end{equation}
where $S_{1,n}(\tau)$ and $S_{2,n}(\tau)$ are defined as follows
\begin{equation}
\label{Gs}
S_{1,n}(\tau)=\dfrac{\ds \int_0^{\bar{l}} \widehat{s}_1(y,\tau) \,  f_{1,n} (y) \, dy}{\ds \int_0^{\bar{l}} f^2_{1,n} (y) \, dy}, \qquad
S_{2,n}(\tau)=\dfrac{\ds \int_{\bar{l}}^1 \widehat{s}_2(y,\tau) \,  f_{2,n} (y) \, dy}{\ds \int_{\bar{l}}^1 f^2_{2,n} (y) \, dy}.
\end{equation}

By substituting the expressions \eqref{SolNoHomegeneo2}-\eqref{Gs} into equation \eqref{Sist_w1}, we obtain the following countable set of homogeneous ordinary differential equations:
\begin{equation}
\label{EDO}
\begin{cases} 
\ds \sum_{n=1}^{\infty} \left[\bar{A_n}'(\tau) +(\bar{\alpha} \, \omega^2_{1,n} - \psi_1) \, \bar{A_n}(\tau) -S_{1,n}(\tau)\right] =0, \\
\ds \sum_{n=1}^{\infty} \left[\bar{B_n}'(\tau) +(\omega^2_{2,n} - \psi_2) \, \bar{B_n}(\tau) -S_{2,n}(\tau)\right] =0,
\end{cases}
\end{equation}
since expansions in eigenfunctions for linear system problems share properties with Fourier series, for the series given in \eqref{EDO} to sum to zero, each term must be zero. This can be addressed through direct integration, leading to:
\begin{equation}
\label{solEDO}
\begin{cases} 
\bar{A_n}(\tau)=\exp\left((\psi_1-\bar{\alpha} \, \omega^2_{1,n})\, \tau\right) \left[K_n + \ds \int_0^\tau   S_{1,n} (s) \, \exp\left((\bar{\alpha} \, \omega^2_{1,n}-\psi_1)\, s\right)  \,ds\right] , \\
\bar{B_n}(\tau)=\exp\left((\psi_2- \omega^2_{2,n})\, \tau\right) \left[K_n + \ds \int_0^\tau   S_{2,n} (s) \, \exp\left((\omega^2_{2,n}-\psi_2)\, s\right)  \,ds\right] ,
\end{cases}
\end{equation}
Only $K_n$ remains to be determined. This sequence can be found by imposing the initial conditions of \eqref{Sist_w1} and using the orthogonality condition, which will be detailed in Section \ref{Ortogonalidad}. Thus, we obtain:
\begin{equation}
\label{CI}
K_n= \dfrac{\dfrac{\varphi \, \phi-\eta \, \mu}{\bar{\alpha}} \, \ds \int_0^{\bar{l}} \Theta_{1,0}(y)\, f_{1,n}(y) \, dy +\ds \int_{\bar{l}}^1 \Theta_{2,0}(y)\, f_{2,n}(y) \, dy }{\dfrac{\varphi \, \phi-\eta \, \mu}{\bar{\alpha}} \ds \int_0^{\bar{l}}  [f_{1,n}(y)]^2 \, dy + \ds \int_{\bar{l}}^1  [f_{2,n}(y)]^2 \, dy}.
\end{equation}
\section{Study of eigenvalues} \label{Autovalores}

The spatial eigenvalues $\omega_{1,n}$ and $\omega_{2,n}$ directly depend on the temporal eigenvalues $\lambda_n$. The solution to the problem of interest can be expressed as an infinite series based on the principle of superposition. This principle assumes that the solution set of the transcendental eigenvalue equation is countably infinite, meaning there are infinitely many solutions $\lambda_n$ that satisfy the eigenvalue equation.

In this work, only real eigenvalues will be considered, as we assume that neither overheating nor thermal runaway occurs in the process under study. In the case of thermal runaway, imaginary eigenvalues may appear, as discussed in \cite{Jain21}. 

This section discusses the existence of infinitely many real solutions $\lambda_n$ to the eigenvalue equation, which is given by:
\begin{equation}
\label{Tangbis}
\tan (\omega_{2,n})=\dfrac{\omega_{2,n} \, B_n- \bar{{Bi}}_2 \, A_n}{\bar{{Bi}}_2 \, B_n+\omega_{2,n}\,  A_n},
\end{equation}
where
\begin{equation}
\label{An2}
\begin{split}
A_n=& \dfrac{\sin(\omega_{1,n} \, \bar{l})}{\cos(\omega_{2,n} \, \bar{l})} \left(\phi \, \dfrac{\bar{{Bi}}_1}{\omega_{1,n}}-\mu \, \omega_{1,n} \right)
  + \dfrac{\cos(\omega_{1,n} \, \bar{l})}{\cos(\omega_{2,n} \, \bar{l})} \left(\phi +\mu \, \bar{{Bi}}_1 \right) - \tan (\omega_{2,n} \, \bar{l}) \, B_n,   
\end{split}
\end{equation}
\begin{equation}
\label{Bn2}
\begin{split}
 B_n = & \sin(\omega_{2,n} \, \bar{l}) \left[\sin(\omega_{1,n} \, \bar{l}) \left(\phi \, \dfrac{\bar{{Bi}}_1}{\omega_{1,n}} - \mu \, \omega_{1,n} \,\right) + \cos(\omega_{1,n} \, \bar{l}) \left(\phi + \mu \, \bar{{Bi}}_1 \, \right) \right] \\
 +& \dfrac{\cos(\omega_{2,n} \, \bar{l})}{\omega_{2,n}}  \left[\sin(\omega_{1,n} \, \bar{l})\left(\eta \,\dfrac{ \bar{{Bi}}_1}{\omega_{1,n}} - \varphi \, \omega_{1,n}\right) + \cos(\omega_{1,n} \, \bar{l}) \left(\eta+\varphi \, \bar{{Bi}}_1\right)\right],
\end{split}
\end{equation}
with
\begin{equation}
\label{Omegasbis}
\begin{cases} 
\omega_{1,n}=\omega_{1,n}(x)=\sqrt{\dfrac{x^2+\psi_1}{\bar{\alpha}}}=\sqrt{\dfrac{x^2+\bar{\nu}_1-\bar{\alpha} \chi_1^2}{\bar{\alpha}}}=\sqrt{\dfrac{x^2+\bar{\nu}_1- \frac{Pe_1^2}{4\bar{\alpha}}}{\bar{\alpha}}}
, \vspace{0.1cm} \\ 
\omega_{2,n}=\omega_{2,n}(x)=\sqrt{x^2+\psi_2}=\sqrt{x^2+\bar{\nu}_2- \chi_2^2}    =\sqrt{x^2+\bar{\nu}_2- \frac{Pe_2^2}{4}}
\end{cases}
\end{equation}
and
\begin{equation}
\label{Aux12}
\begin{cases} 
\psi_1= \bar{\nu}_1 - \bar{\alpha} \, \chi_1^2, \quad \psi_2= \bar{\nu}_2 - \chi_2^2, 
 \quad \bar{{Bi}}_1 = {{Bi}_1}^*-\chi_1, \quad \bar{{Bi}}_2 = {{Bi}_2}^*-\chi_2,
 \\ \phi= \xi \,\delta , \quad \mu= \xi\,\bar{R} ,\quad \eta=\xi \, \left(\gamma+ \sigma \,\chi_1 - \delta \,\chi_2   \right), \quad 
\varphi=\xi \, \left(\sigma - \bar{R} \,\chi_2 \right), 
\\ \xi= \exp\left(\, \bar{l} \, (\chi_1-\chi_2)\right), \quad  \delta=1+ \bar{R} \, \chi_1,  \quad
\chi_1= \dfrac{Pe_1}{2 \,\bar{\alpha}}, \,\,\, \chi_2= \dfrac{Pe_2}{2}, 
\\ {{Bi}_1}^*=\dfrac{Pe_2}{\bar{\alpha}}+\dfrac{{Bi}_1}{\bar{\kappa}},  \quad
{{Bi}_2}^*= Pe_2 - {Bi}_2, \quad
\gamma= Pe_2- Pe_1 \, \dfrac{\bar{\kappa}}{\bar{\alpha}},
\\ \sigma=\bar{\kappa} + \bar{R}\,Pe_2, \quad \bar{l}=\dfrac{l}{L}, \quad \bar{R}=\dfrac{R}{L}, \quad \bar{\alpha}=\dfrac{\alpha_1}{\alpha_2}, \quad {Pe}_m=\dfrac{L}{\alpha_2} \, \beta_m,   
\\ \bar{\nu}_m=\dfrac{L^2}{\alpha_2} \,\nu_m, \quad \bar{\kappa}=\dfrac{\kappa_1}{\kappa_2}, \quad {Bi}_m=\dfrac{L}{\kappa_2} \, h_m.
\end{cases}
\end{equation}

Since the equation \eqref{Tangbis}-\eqref{Aux12} is transcendental, it is not possible to obtain explicit solutions. Additionally, analytically proving that this equation has infinitely many solutions in the general case is a difficult task due to its complexity. However, the existence of such solutions can be verified numerically for each specific case.

If we denote by:
\begin{equation}
\label{OtroAuxbis}
r(x)=\dfrac{\omega_{2,n}(x) \, B_n(x)- \bar{{Bi}}_2 \, A_n(x)}{\bar{{Bi}}_2 \, B_n(x)+\omega_{2,n}(x)\,  A_n(x)}, \qquad 
q(x)=\tan (\omega_{2,n}(x)), \
\end{equation}
showing that the eigenvalue equation has infinitely many real solutions reduces to seeing that the functions $r(x)$ and $q(x)$ intersect infinitely many times. As an example, we will observe this in two specific cases, where the thermal parameters were taken from \cite{Cengel07}, and the other physical parameters are selected to ensure that the heat transfer problem is meaningful.
\setcounter{case}{0}
\setcounter{table}{0}
\begin{case}
\label{case1}
The heat transfer problem in a $Fe-Al$ bilayer body, is considered. The parameters used are listed in the following table \ref{PPC1}:
\end{case}

\begin{case}
\label{case2}
The heat transfer problem in a $Cu-Pb$ bilayer body, is considered. The parameters used are listed in the following table \ref{PPC2}:
\end{case}

\begin{figure}[!h]
\begin{center}
\includegraphics[width=0.495\textwidth]{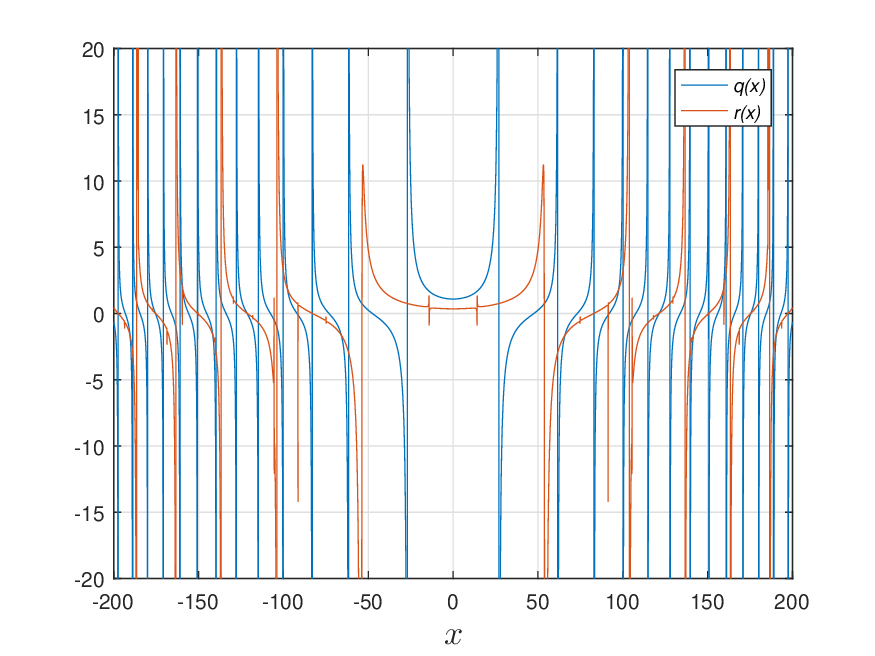}
\includegraphics[width=0.495\textwidth]{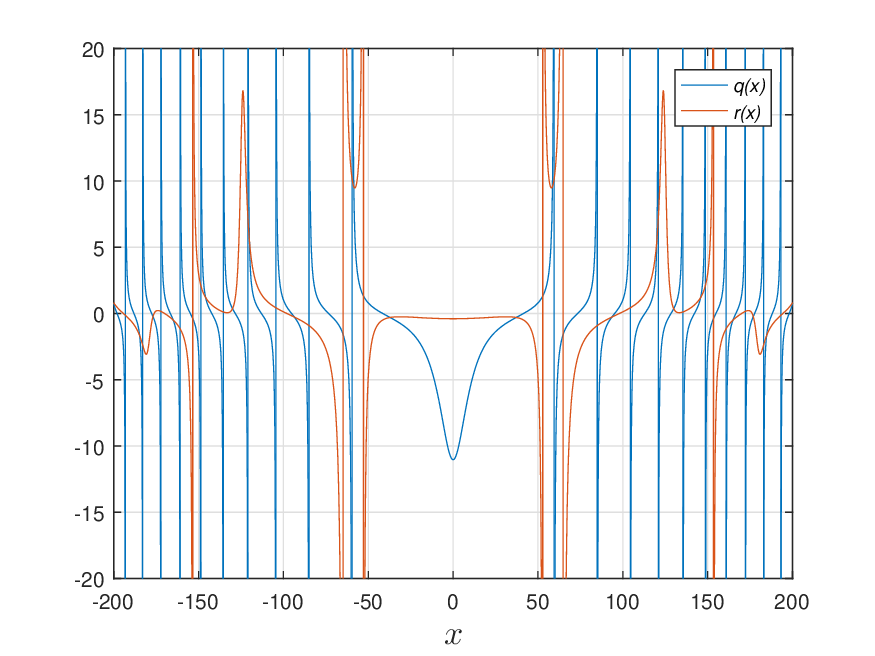}
\vspace{-0.5cm} 
\caption{Scheme of intersection of the functions $q(x)$ and $r(x)$. On the left, for case \ref{case1}, and on the right, for case \ref{case2}.}
\vspace{-0.5cm}
\label{inf_sol}
\end{center}
\end{figure}

From Fig.\ref{inf_sol} the intercessions for $\lambda_n \in (-200,200)$, can be observed. It can be inferred, for both cases, that the functions $q(x)$ and $r(x)$ intersect infinitely many times.

\begin{table}[h!]
\begin{center}
\begin{minipage}{0.45\textwidth}
\centering
\begin{tabular}{lc} \toprule
Parameters                                                       &       Values   \\ \midrule
$\alpha_1 \left(\times 10^{4}\right) \, \left[m^{2}/s\right]$  &       0.20451    \\ 
$\alpha_2 \left(\times 10^{4}\right) \, \left[m^{2}/s\right]$  &       0.84010   \\ 
$\kappa_1 \, \left[W/m^{\circ}C\right] $                         &       73        \\ 
$\kappa_2 \, \left[W/m^{\circ}C\right] $                         &       204       \\ 
$L \, \left[m\right] $                                           &       1       \\ 
$l \, \left[m\right] $                                           &       0.4      \\ 
$h_1 \, \left[W/m^2\,^{\circ}C\right] $                            &       12        \\   
$h_2 \, \left[W/m^2\,^{\circ}C\right] $                            &       10        \\    
$\beta_1 \, \left[m/s\right] $                                     &       0.001       \\   
$\beta_2 \, \left[m/s\right] $                                     &       0.002       \\ 
$\nu_1 \, \left[1/s\right] $                                     &       10       \\   
$\nu_2 \, \left[1/s\right] $                                     &       20       \\ 
$R\, \left[m\right] $                                                  &       0.05        \\   \bottomrule
\end{tabular}
\caption{Physical parameters of case \ref{case1}.}
\vspace{-0.5cm}
\label{PPC1}
\end{minipage}
\hspace{0.05\textwidth}
\begin{minipage}{0.45\textwidth}
\centering
\begin{tabular}{lc} \toprule
Parameters                                                       &       Values   \\ \midrule
$\alpha_1 \left(\times 10^{4}\right) \, \left[m^{2}/s\right]$  &       1.12530    \\ 
$\alpha_2 \left(\times 10^{4}\right) \, \left[m^{2}/s\right]$  &       0.23673   \\ 
$\kappa_1 \, \left[W/m^{\circ}C\right] $                         &       386        \\ 
$\kappa_2 \, \left[W/m^{\circ}C\right] $                         &       35       \\ 
$L \, \left[m\right] $                                           &       2       \\ 
$l \, \left[m\right] $                                           &       0.8      \\ 
$h_1 \, \left[W/m^2\,^{\circ}C\right] $                            &       10        \\   
$h_2 \, \left[W/m^2\,^{\circ}C\right] $                            &       12        \\ 
$\beta_1 \, \left[m/s\right] $                                     &       0.002       \\   
$\beta_2 \, \left[m/s\right] $                                     &       0.001       \\ 
$\nu_1 \, \left[1/s\right] $                                     &       1       \\   
$\nu_2 \, \left[1/s\right] $                                     &       2       \\    
$R\, \left[m\right] $                                                  &       0.07        \\   \bottomrule
\end{tabular}
\caption{Physical parameters of case \ref{case2}.}
\vspace{-0.5cm}
\label{PPC2}
\end{minipage}
\end{center}
\end{table}

\section{Study of the orthogonality relationship} \label{Ortogonalidad}

In this section, we will derive the orthogonality condition, or principle, for this problem. This result is necessary to determine the sequence $K_n$ in \eqref{CI}. As shown in \eqref{fs} for two indices $n$ and $j$ the functions $f_{1,n}$, $f_{1,j}$, $f_{2,n}$ and $f_{2, j}$ must satisfy the following:
\begin{equation}
\label{f1nj}
\begin{cases} 
\bar{\alpha} \, f''_{1,n}(y)+\psi_1 \,f_{1,n}(y)= - \lambda_n^2 \,f_{1,n}(y), \quad & y \in (0,\bar{l}), \\ 
\bar{\alpha} \, f''_{1,j}(y)+\psi_1 \,f_{1,j}(y)= - \lambda_j^2 \,f_{1,j}(y), \quad & y \in (0,\bar{l})
\end{cases}
\end{equation}
and
\begin{equation}
\label{f2nj}
\begin{cases} 
 f''_{2,n}(y)+\psi_2 \,f_{2,n}(y)= - \lambda_n^2 \,f_{2,n}(y), \quad & y \in (\bar{l},1), \\ 
 f''_{2,j}(y)+\psi_2 \,f_{2,j}(y)= - \lambda_j^2 \,f_{2,j}(y), \quad & y \in (\bar{l},1).
\end{cases}
\end{equation}
Multiply the first equation of \eqref{f1nj} by $f_{1,j}$ and the second by $f_{1,n}$.
Similarly, multiply the first equation of \eqref{f2nj} by $f_{2,j}$
and the second by $f_{2,n}$. This gives rise to the following expressions:
\begin{equation}
\label{f1njbis}
\begin{cases} 
\bar{\alpha} \, f''_{1,n}(y) \,f_{1,j}(y)+\psi_1 \,f_{1,n}(y)\,f_{1,j}(y)= - \lambda_n^2 \,f_{1,n}(y)\,f_{1,j}(y), \quad & y \in (0,\bar{l}),   \\ 
\bar{\alpha} \, f''_{1,j}(y)\,f_{1,n}(y)+\psi_1 \,f_{1,j}(y)\,f_{1,n}(y)= - \lambda_j^2 \,f_{1,j}(y)\,f_{1,n}(y) \quad & y \in (0,\bar{l}), 
\end{cases}
\end{equation}
and
\begin{equation}
\label{f2njbis}
\begin{cases} 
 f''_{2,n}(y) \,f_{2,j}(y)+\psi_2 \,f_{2,n}(y)\,f_{2,j}(y)= - \lambda_n^2 \,f_{2,n}(y)\,f_{2,j}(y), \quad & y \in (\bar{l},1),  \\ 
 f''_{2,j}(y)\,f_{2,n}(y)+\psi_2 \,f_{2,j}(y)\,f_{2,n}(y)= - \lambda_j^2 \,f_{2,j}(y)\,f_{2,n}(y), \quad & y \in (\bar{l},1).
\end{cases}
\end{equation}
Subtract the two expressions in \eqref{f1njbis}, and similarly, do the same for \eqref{f2njbis}, to obtain
\begin{equation}
\label{fnjbis}
\begin{cases} 
\bar{\alpha} \, \left[f''_{1,n}(y) \,f_{1,j}(y)-f''_{1,j}(y)\,f_{1,n}(y)\right]= (\lambda_j^2- \lambda_n^2) \,f_{1,n}(y)\,f_{1,j}(y),  \quad & y \in (0,\bar{l}),\\ 
f''_{2,n}(y) \,f_{2,j}(y)-f''_{2,j}(y)\,f_{2,n}(y)= (\lambda_j^2- \lambda_n^2) \,f_{2,n}(y)\,f_{2,j}(y), \quad & y \in (\bar{l},1),
\end{cases}
\end{equation}
the equations in \eqref{fnjbis} are conveniently rewritten, and the first equation is then multiplied by $\dfrac{\varphi \,\phi-\eta \,\mu}{\bar{\alpha}}$.
\begin{equation}
\label{fnjbis2}
\begin{cases} 
(\varphi \,\phi-\eta \,\mu ) \left[f'_{1,n}(y) \,f_{1,j}(y)-f'_{1,j}(y)\,f_{1,n}(y)\right]'= \dfrac{\varphi \,\phi-\eta \,\mu}{\bar{\alpha}} (\lambda_j^2- \lambda_n^2) f_{1,n}(y)f_{1,j}(y), \\ 
\left[f'_{2,n}(y) \,f_{2,j}(y)-f'_{2,j}(y)\,f_{2,n}(y)\right]'= (\lambda_j^2- \lambda_n^2) \,f_{2,n}(y)\,f_{2,j}(y), 
\end{cases}
\end{equation}
the equalities are integrated over their respective intervals of definition and then added. This yields,
\begin{equation}
\label{fnjbis3}
\begin{split}
&(\lambda_j^2- \lambda_n^2) \left\{\dfrac{\varphi \,\phi-\eta \,\mu}{\bar{\alpha}} \ds \int_0^{\bar{l}}  f_{1,n}(y)f_{1,j}(y) \, dy + \ds \int_{\bar{l}}^1  f_{2,n}(y)f_{2,j}(y) \, dy \right\} \\
= & (\varphi \,\phi-\eta \,\mu ) \left[f'_{1,n}(y) \,f_{1,j}(y)-f'_{1,j}(y)\,f_{1,n}(y)\right] \Big|_{0}^{\bar{l}} \\
 & +\left[f'_{2,n}(y) \,f_{2,j}(y)-f'_{2,j}(y)\,f_{2,n}(y)\right] \Big|_{\bar{l}}^1,
\end{split}
\end{equation}
after performing operations on the term on the right-hand side and applying the boundary conditions from \eqref{fs}, we obtain:
\begin{equation}
\label{fnjbis4}
\begin{split}
&(\lambda_j^2- \lambda_n^2) \left\{\dfrac{\varphi \,\phi-\eta \,\mu}{\bar{\alpha}} \ds \int_0^{\bar{l}}  f_{1,n}(y)f_{1,j}(y) \, dy + \ds \int_{\bar{l}}^1  f_{2,n}(y)f_{2,j}(y) \, dy \right\} \\
= & (\varphi \,\phi-\eta \,\mu ) \left[f'_{1,n}(\bar{l}) \,f_{1,j}(\bar{l})-f'_{1,j}(\bar{l})\,f_{1,n}(\bar{l})\right] + \left[f'_{2,j}(\bar{l})\,f_{2,n}(\bar{l})-f'_{2,n}(\bar{l}) \,f_{2,j}(\bar{l})\right] ,
\end{split}
\end{equation}
From the interface conditions of \eqref{fs}, it follows

\begin{equation}
\label{fnjbis5}
\begin{split}
&(\lambda_j^2- \lambda_n^2) \left\{\dfrac{\varphi \,\phi-\eta \,\mu}{\bar{\alpha}} \ds \int_0^{\bar{l}}  f_{1,n}(y)f_{1,j}(y) \, dy + \ds \int_{\bar{l}}^1  f_{2,n}(y)f_{2,j}(y) \, dy \right\} \\
= & (\varphi \,\phi-\eta \,\mu ) \left[f'_{1,n}(\bar{l}) \,f_{1,j}(\bar{l})-f'_{1,j}(\bar{l})\,f_{1,n}(\bar{l})\right] 
\\- & 
\left[(\eta \, f_{1,n}(\bar{l}) + \varphi \, f'_{1,n}(\bar{l})) \, (\phi \, f_{1,j}  (\bar{l}) +\mu \, f'_{1,j}(\bar{l})) \right] 
\\+ & (\eta \, f_{1,j}(\bar{l}) + \varphi \, f'_{1,j}(\bar{l})) \, (\phi \, f_{1,n}  (\bar{l}) +\mu \, f'_{1,n}(\bar{l})).
\end{split}
\end{equation}
By performing algebraic operations, the orthogonality condition is obtained. For $n \neq j$ it holds that
\begin{equation}
\label{CondOrtogonalidad}
(\lambda_j^2- \lambda_n^2) \left\{\dfrac{\varphi \,\phi-\eta \,\mu}{\bar{\alpha}} \ds \int_0^{\bar{l}}  f_{1,n}(y)f_{1,j}(y) \, dy + \ds \int_{\bar{l}}^1  f_{2,n}(y)f_{2,j}(y) \, dy \right\} =0.
\end{equation}

\section{Consistency of the solution} \label{Consistencia}

There are several ways to analyze the consistency of the solution obtained with those found in the literature. In \cite{Jain21} the authors consider a situation similar to the one addressed here, but with simpler characteristics that are of special interest for this analysis. In that article, external heat sources are neglected, and thermal contact resistance at the interface is not considered. We aim to verify that under these assumptions, both solutions are equivalent.

In this case, with no external heat sources, we have $s_1=s_2=0$. Furthermore, since the thermal resistance at the interface is neglected, $R=0$.

Given that the external sources are zero $(s_1=s_2=0)$, the problem reduces to solving the associated homogeneous system given by:
\begin{equation}
\label{Sol2}
\begin{cases} 
\Theta_1^H(y,\tau)= \ds \sum_{n=1}^{\infty} { K_n \left[A_{1,n}  \cos (\omega_{1,n} \, y) + B_{1,n}  \sin (\omega_{1,n}\, y) \right] \exp(-\lambda_n^2 \, \tau)},  \\ 
\Theta_2^H(y,\tau)= \ds \sum_{n=1}^{\infty} {K_n \left[A_{2,n}  \cos (\omega_{2,n} \,y) + B_{2,n}  \sin (\omega_{2,n} \, y) \right] \exp(-\lambda_n^2 \, \tau)},
\end{cases}
\end{equation}
where 
\begin{equation}
\label{Omegas2}
\begin{cases} 
\omega_{1,n}=\omega_{1,n}(\lambda_n)=\sqrt{\dfrac{\lambda_n^2+\psi_1}{\bar{\alpha}}}=\sqrt{\dfrac{\lambda_n^2+\bar{\nu}_1-\bar{\alpha} \chi_1^2}{\bar{\alpha}}}=\sqrt{\dfrac{\lambda_n^2+\bar{\nu}_1- \frac{Pe_1^2}{4\bar{\alpha}}}{\bar{\alpha}}}
, \vspace{0.1cm} \\ 
\omega_{2,n}=\omega_{2,n}(\lambda_n)=\sqrt{\lambda_n^2+\psi_2}=\sqrt{\lambda_n^2+\bar{\nu}_2- \chi_2^2}    =\sqrt{\lambda_n^2+\bar{\nu}_2- \frac{Pe_2^2}{4}}.
\end{cases}
\end{equation}
wich
$A_{1,n}=1$, $B_{1,n}=\frac{\bar{{Bi}}_1}{\omega_{1,n}}$,
$A_{2,n}=A_n$ and $B_{2,n}=B_n$ where
\begin{equation}
\label{An2bis}
\begin{split}
A_n=& \dfrac{\sin(\omega_{1,n} \, \bar{l})}{\cos(\omega_{2,n} \, \bar{l})} \left(\phi \, \dfrac{\bar{{Bi}}_1}{\omega_{1,n}}-\mu \, \omega_{1,n} \right)
  + \dfrac{\cos(\omega_{1,n} \, \bar{l})}{\cos(\omega_{2,n} \, \bar{l})} \left(\phi +\mu \, \bar{{Bi}}_1 \right) - \tan (\omega_{2,n} \, \bar{l}) \, B_n   
\end{split}
\end{equation}
and
\begin{equation}
\label{Bn2bis}
\begin{split}
 B_n = & \sin(\omega_{2,n} \, \bar{l}) \left[\sin(\omega_{1,n} \, \bar{l}) \left(\phi \, \dfrac{\bar{{Bi}}_1}{\omega_{1,n}} - \mu \, \omega_{1,n} \,\right) + \cos(\omega_{1,n} \, \bar{l}) \left(\phi + \mu \, \bar{{Bi}}_1 \, \right) \right] \\
 +& \dfrac{\cos(\omega_{2,n} \, \bar{l})}{\omega_{2,n}}  \left[\sin(\omega_{1,n} \, \bar{l})\left(\eta \,\dfrac{ \bar{{Bi}}_1}{\omega_{1,n}} - \varphi \, \omega_{1,n}\right) + \cos(\omega_{1,n} \, \bar{l}) \left(\eta+\varphi \, \bar{{Bi}}_1\right)\right].
\end{split}
\end{equation}
The eigenvalues $\lambda_n$ are the infinitily many solutions of the equation
\begin{equation}
\label{Tang2}
\tan (\omega_{2,n})=\dfrac{\omega_{2,n} \, B_n- \bar{{Bi}}_2 \, A_n}{\bar{{Bi}}_2 \, B_n+\omega_{2,n}\,  A_n}.
\end{equation}
Finally, $K_n$ is determined from the initial conditions using the orthogonality principle discussed in section \ref{Ortogonalidad}.
\begin{equation}
\label{CI2}
K_n= \dfrac{\dfrac{\varphi \, \phi-\eta \, \mu}{\bar{\alpha}} \, \ds \int_0^{\bar{l}} \Theta_{1,0}(y)\, f_{1,n}(y) \, dy +\ds \int_{\bar{l}}^1 \Theta_{2,0}(y)\, f_{2,n}(y) \, dy }{\dfrac{\varphi \, \phi-\eta \, \mu}{\bar{\alpha}} \ds \int_0^{\bar{l}}  [f_{1,n}(y)]^2 \, dy + \ds \int_{\bar{l}}^1  [f_{2,n}(y)]^2 \, dy}.
\end{equation}
The only remaining step is to impose the absence of contact resistance at the interface. To achieve this, we need to evaluate the equations \eqref{Sol2}-\eqref{CI2} with $R=0$.
 The parameters affected by these changes are:
\begin{equation}
\label{Aux1bis2}
\phi= \xi , \qquad \mu= 0 ,\qquad  \varphi=\xi \,\bar{\kappa}  \qquad  \eta=\xi \, \left(\gamma+ \bar{\kappa}\,\chi_1 - \chi_2   \right).
\end{equation}
In summary, when we examine the solution (derived in this article) for the specific case of transient heat transfer with no thermal sources and neglecting contact resistance at the interface, it is found that the solution satisfies the conditions provided by the authors in \cite{Jain21}.

\section{Numerical Modelling}

The analytical solution of such problems involves a significant numerical burden, making it complex to obtain temperature profiles for specific cases. Consequently, these problems are often modeled using numerical methods that facilitate graphing different temperature profiles and extracting relevant information.

The finite difference method is frequently employed for evolutionary heat transfer problems. In the context of multilayer bodies, handling the junction between different materials can be challenging, especially when temperature continuity is not maintained. Some authors have addressed this issue by introducing virtual or artificial layers; see, for example, \cite{Yuan22}.

In this work, we propose an explicit second-order finite difference method, using a forward scheme in time and a centered scheme in space, with specific adaptations at the boundaries and interface. At the right boundary, we apply backward differences, while at the left boundary, we use forward differences. For the interface, the approach involves forward or backward differences depending on whether the material to the left or right is considered. Specifically, the first-order discretization at the interface is given by:
\begin{equation}
\label{Disct_Interf}
\dfrac{\partial T_1}{\partial x}(l,t) = \dfrac{1}{\Delta x}\left(T^{1}_{n_l,j} - T^{1}_{n_l-1,j}\right), \qquad \dfrac{\partial T_2}{\partial x}(l,t) = \dfrac{1}{\Delta x}\left(T^{2}_{n_l+1,j} - T^{2}_{n_l,j}\right).
\end{equation}

In order to implement the numerical method, two uniform 2D partitions are defined in the spatial variable $x$ and the time variable $t$ as a discrete set $\mathcal{P}$ that satisfies:
\begin{equation}
\label{particion1}
\begin{cases}
\mathcal{P}_1=\{ (x_i,t_j)/ \, i=1,2,...,n_l ;\, j=1,2,...,J;\, x_i \in \mathcal{P}_x^1,\, t_j \in \mathcal{P}_t \}, \\
\mathcal{P}_2=\{ (x_i,t_j)/ \, i=n_l,n_{l+1},...,n_L ;\, j=1,2,...,J;\, x_i \in \mathcal{P}_x^2,\, t_j \in \mathcal{P}_t \},
\end{cases}
\end{equation}
where
\begin{equation}
\label{particion2}
\begin{cases}
\mathcal{P}_x^1=\{ x_1< \cdots < x_i< \cdots <x_{n_l}, \,\,\, x_i=(i-1) \Delta x, \,\,\, i=1,2,...,n_l\} \\
\mathcal{P}_x^2=\{ x_{n_l}<  \cdots < x_i< \cdots <x_{n_L}, \,\,\, x_i=(i-1) \Delta x, \,\,\, i=n_l, n_{l+1},...,n_L\}
\end{cases}
\end{equation}
and
\begin{equation}
\label{particion3}
\mathcal{P}_t=\{ t_1< t_2< \cdots < t_j< \cdots <t_M, \,\,\, t_j=(j-1) \Delta t, \,\,\, j=1, 2,...,J\}.
\end{equation}

Specifically, $\mathcal{P}_x^i$ with $i=1,2$ denotes the partition of the spatial variable $x$ associated with $T_i$, while $\mathcal{P}_t$ denotes the corresponding partition associated with the time variable $t$. The values of $\Delta x$ and $\Delta t$ correspond to the spatial and temporal discretization steps, respectively. These values are numerically determined and defined on an equidistant (uniform) grid as $\Delta x = x_i-x_{i-1}$ and $\Delta t = t_j-t_{j-1}$.

To find the numerical solution to the heat transfer problem under study, equations \eqref{Ec_Parab}-\eqref{Cond_Inicial} are discretized according to this scheme. Consequently, the following algebraic system can be derived:

\begin{equation}
\label{Discret}
\begin{cases} 
T^1_{i,j+1}=\zeta_{11}\, T^1_{i+1,j}+\zeta_{12}\, T^1_{i,j} + \zeta_{13}\, T^1_{i-1,j}+ P^1_{i,j}, &  i=2,...,n_{l-1},  j=2,...,J,\\
T^2_{i,j+1}=\zeta_{21}\, T^2_{i+1,j}+\zeta_{22}\, T^2_{i,j} + \zeta_{23}\, T^2_{i-1,j}+ P^2_{i,j}, &  i=n_{l+1},...,n_{L-1},  j=2,...,J,\\
T^1_{i,j}=T^1_i,  &  i=1,2,...,n_l, \,\, j=1, \\
T^2_{i,j}=T^2_i,  &  i=n_l,...,n_L, \,\, j=1, \\
T^1_{i,j}=\epsilon_1 \, T^1_{i+1,j} ,  &  i=1, \, j=2,...,J,\\ 
T^2_{i,j}=\epsilon_2 \, T^2_{i-1,j},  &  i=n_L, \, j=2,...,J, \\
T^1_{i,j}=\upsilon_{11} T^1_{i-1,j}+ \upsilon_{12} T^2_{i+1,j},  &  i=n_l, \, j=2,...,J,\\
T^2_{i,j}=\upsilon_{21} T^1_{i-1,j}+ \upsilon_{22} T^2_{i+1,j},  &  i=n_l, \, j=2,...,J,\\
\end{cases}
\end{equation}
where

\begin{equation}
\label{OtroAux}
\begin{cases} 
\zeta_{11}=\dfrac{\alpha_1 \,\Delta t }{(\Delta x)^2}-\dfrac{\beta_1 \,\Delta t }{2 \, \Delta x}, \quad
\zeta_{12}=1- 2 \dfrac{\alpha_1 \,\Delta t }{(\Delta x)^2}+\nu_1 \, \Delta t, \quad
\zeta_{13}=\dfrac{\alpha_1 \,\Delta t }{(\Delta x)^2}+\dfrac{\beta_1 \,\Delta t }{2 \, \Delta x},  \vspace{0.2cm} \\
\zeta_{21}=\dfrac{\alpha_2 \,\Delta t }{(\Delta x)^2}-\dfrac{\beta_2 \,\Delta t }{2 \, \Delta x}, \quad
\zeta_{22}=1- 2 \dfrac{\alpha_2 \,\Delta t }{(\Delta x)^2}+\nu_2 \, \Delta t, \quad
\zeta_{23}=\dfrac{\alpha_2 \,\Delta t }{(\Delta x)^2}+\dfrac{\beta_2 \,\Delta t }{2 \, \Delta x}, \vspace{0.2cm} \\
P^1_{i,j}=s^1_{i,j} \, \Delta t, \quad
P^2_{i,j}=s^2_{i,j} \, \Delta t, \quad 
\epsilon_1=\dfrac{1}{1+\Delta x \, \Pi_1},  \quad  \epsilon_2=\dfrac{1}{1-\Delta x \, \Pi_2 },\vspace{0.2cm} \\
\Pi_1=\dfrac{\beta_1}{\alpha_1}+\dfrac{h_1}{\kappa_1}, \quad \Pi_2= \dfrac{\beta_2}{\alpha_2}-\dfrac{h_2}{\kappa_2}, \quad \upsilon_{11}=\dfrac{\kappa_1 + \kappa_2 \, \Omega  Z_2}{\Lambda}, \quad \upsilon_{12}=\dfrac{\kappa_2}{\Lambda},\vspace{0.2cm} \\
\upsilon_{21}=\dfrac{\kappa_1 \, (\Omega+1) - \kappa_1 \, \Omega \, Z_1}{\Lambda}, \quad \upsilon_{22}=\dfrac{\kappa_2 \, (\Omega+1) }{\Lambda}, \quad \Lambda= \kappa_1  \, Z_1 + \kappa_2  \, Z_2 \, (\Omega+1), \vspace{0.2cm} \\
\Omega=\dfrac{R}{\Delta x}, \quad Z_1= 1- \dfrac{\beta_1}{\alpha_1} \, \Delta x, \quad Z_2= 1+ \dfrac{\beta_2}{\alpha_2} \, \Delta x.
\end{cases}
\end{equation}

The convergence and stability conditions of this method are documented in the literature \cite{Morton05}. For the problem addressed here, these conditions take the form:
\begin{equation}
\label{cond_est}
\left(\dfrac{\beta_1 \,\Delta t }{2 \, \Delta x}\right)^2<2 \, \dfrac{\alpha_1 \,\Delta t }{(\Delta x)^2}< 1, \qquad
\left(\dfrac{\beta_2 \,\Delta t }{2 \, \Delta x}\right)^2<2 \, \dfrac{\alpha_2 \,\Delta t }{(\Delta x)^2}< 1.
\end{equation}

Under these conditions, it is guaranted a precision of first order in time and of second order in space for the algebraic problem \eqref{Discret}-\eqref{OtroAux}.

\section{Numerical Example}

To perform the numerical example, a non-parallel computational scheme is implemented in MATLAB. The simulated results are obtained within few minutes using an Intel(R) Core(TM) i7-6700K 4 GHz processor. In all cases, it is assumed that the dissipative fluid is air at normal pressure. The convective heat transfer coefficients $h_1$ and $h_2$ are determined according to \cite{Umbricht20conv}, and the thermal parameters of the materials are taken from \cite{Cengel07} and summarized in the following table.

\setcounter{table}{0}
\begin{table}[h!]
\begin{center}
{\begin{tabular}{lccc} \toprule
Materials & Symbol & $\alpha^2 \left(\times 10^{4}\right) \, \left[m^{2}/s\right] $  & $\kappa \, \left[W/m^{\circ}C\right] $\\ \midrule
Lead  &     Pb       &                 0.23673                                  &              35                      \\        
Iron   & Fe         &                 0.20451                                  &              73                      \\ 
Nickel  & Ni         &                  0.22663                                  &              90                      \\ 
Aluminium   & Al         &                 0.84010                                  &              204                     \\ 
Copper    &  Cu         &                 1.12530                                  &              386                     \\ 
Silver   &  Ag         &                 1.70140                                  &              419                     \\ \bottomrule
\end{tabular}}  
\end{center}
\vspace{-0.5cm}
\caption{Thermal properties of different materials.}
\label{Prop_Termicas}
\end{table}
\begin{xmpl}
\label{example1}

For this example, the following parameters are considered:
$L=1 \, m$, $l=0.4 \, m$, $t_\infty=72000 \, s=20 \, h$, $\beta_1=\beta_2=0.02 \, m/s$ , $\nu_1=\nu_2=-0.0003 \,\, 1/s$, $R=0.05 \, m$.

The initial conditions are $T_{1,0}(x)=T_{2,0}(x)=0$ and the heat generation source $s(x,t)$ is a continuous and differentiable function given by:
\begin{equation}
\label{fuente}
\begin{cases}
s_1(x,t)= \dfrac{25}{l \, T^2} \, \dfrac{^{\circ}C}{m \, s} \, x \, (x-l)\, t\, (t-t_\infty), & \, (x,t) \in [0,l] \times [0,t_\infty],  \\
\\
s_2(x,t)=\dfrac{25}{(L-l) \, T^2} \, \dfrac{^{\circ}C}{m \, s} \, (x-l) \, (x-L) \, t \, (t-t_\infty), & \, (x,t) \in [l,L] \times [0,t_\infty].
\end{cases}
\end{equation}
\end{xmpl}
%
\begin{figure}[!h]
\begin{center}
\includegraphics[width=0.495\textwidth]{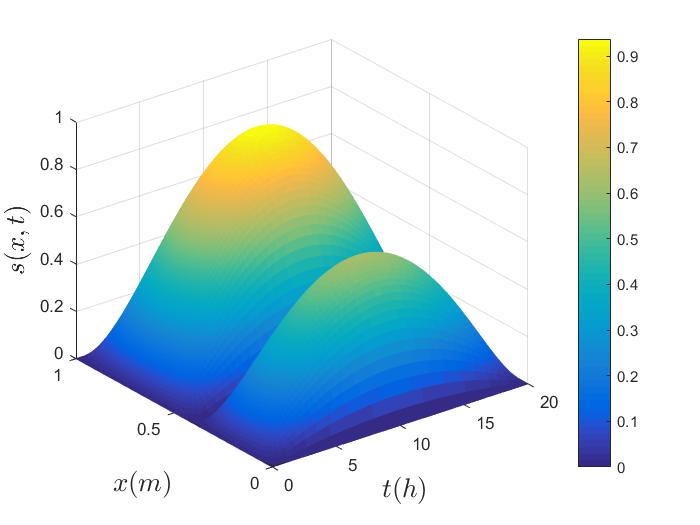}
\includegraphics[width=0.495\textwidth]{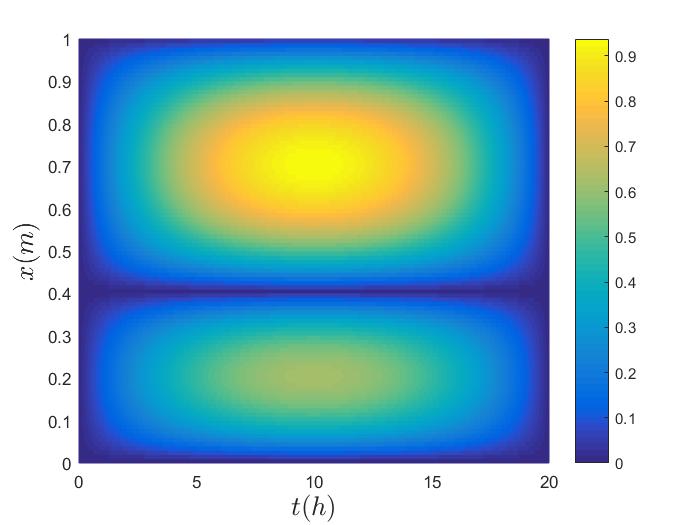}
\vspace{-0.5cm} 
\caption{Heat source.}
\vspace{-0.5cm}
\label{Heat_source}
\end{center}
\end{figure}

The thermal source acting on the body is of fundamental importance as it directly affects the functional form of the temperature profiles. The source given by \eqref{fuente} is an interesting function because it models heating that originates from the center of each layer and decreases towards the edges, where heat generation is zero. In figure \ref{Heat_source}, it can be observed that the maximum heating rate in the first layer is $0.5 \, ^{\circ}C/s$ and in the second layer, it is approximately $0.90 \, ^{\circ}C/s$. These maximum heat generation sources occur at $x=0.2 \, m, t=10 \, h$ for the first layer and at $x=0.7 \, m$, $t=10 \, h$ for the second layer.

\begin{figure}[!h]
\begin{center}
\includegraphics[width=0.495\textwidth]{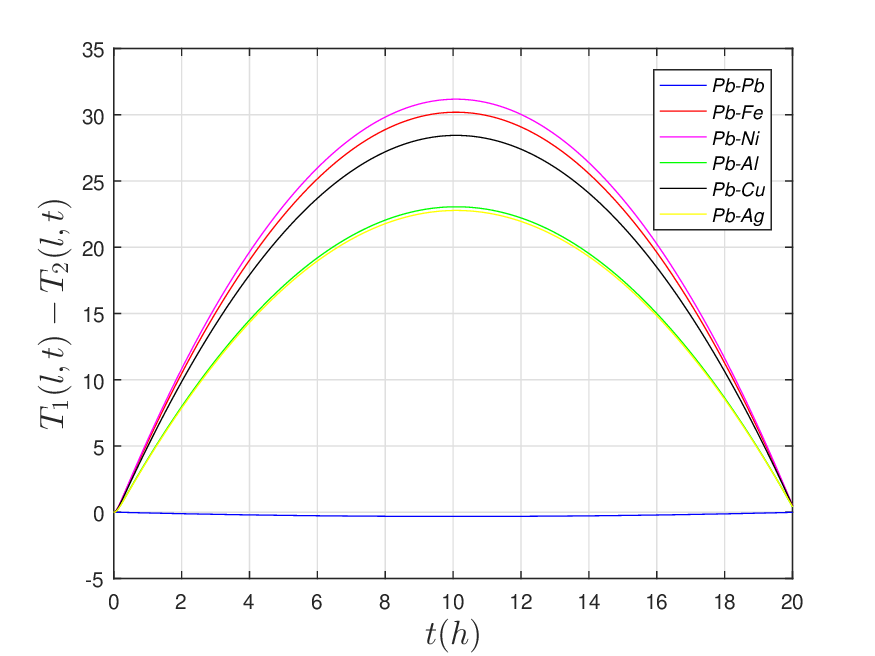}
\includegraphics[width=0.495\textwidth]{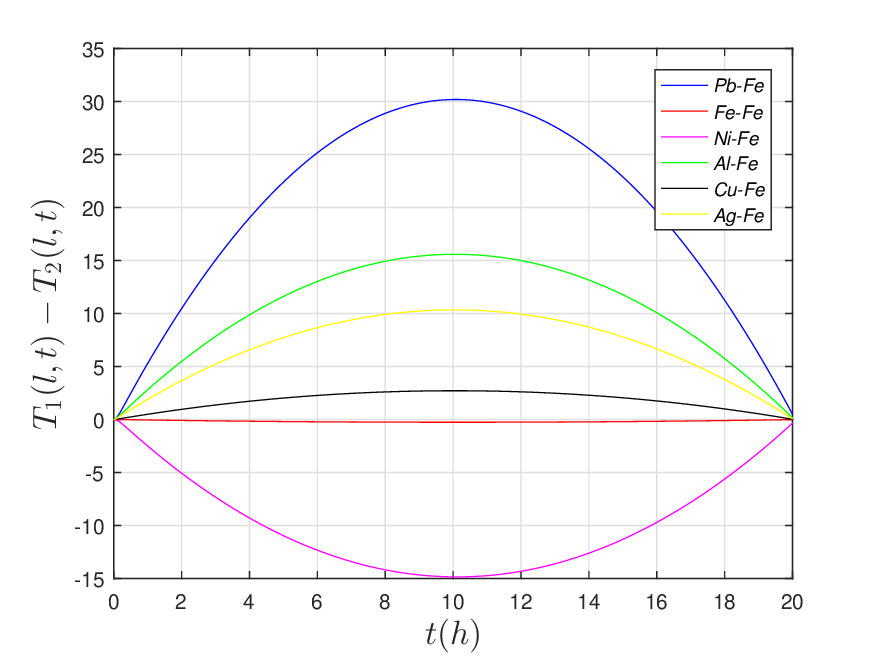}
\vspace{-0.5cm}
\caption{Temperature difference between the layers (interface), $Pb-Material$ (left) and $Material-Fe$ (right) for different materials.}
\vspace{-0.5cm}
\label{Dif_dist_mat}
\end{center}
\end{figure}
Figure \ref{Dif_dist_mat} shows the temperature difference at the interface due to the thermal jump. On the left, the difference profiles for $Pb-Material$ are plotted, while the difference profiles for $Material-Fe$ are shown on the right. Different materials are considered in both cases. Note that in most cases, this difference is positive, which results from the relationship between the thermal diffusivities. Additionally, in all cases, it can be observed that the maximum temperature difference occurs at $t=10 \, h$.
%
\begin{figure}[!h]
\begin{center}
\includegraphics[width=0.495\textwidth]{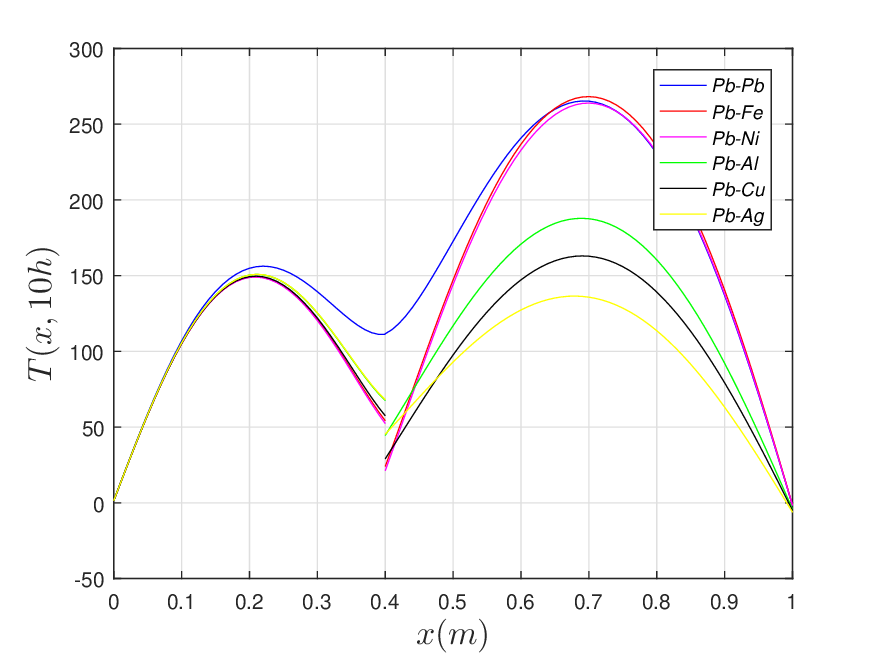}
\includegraphics[width=0.495\textwidth]{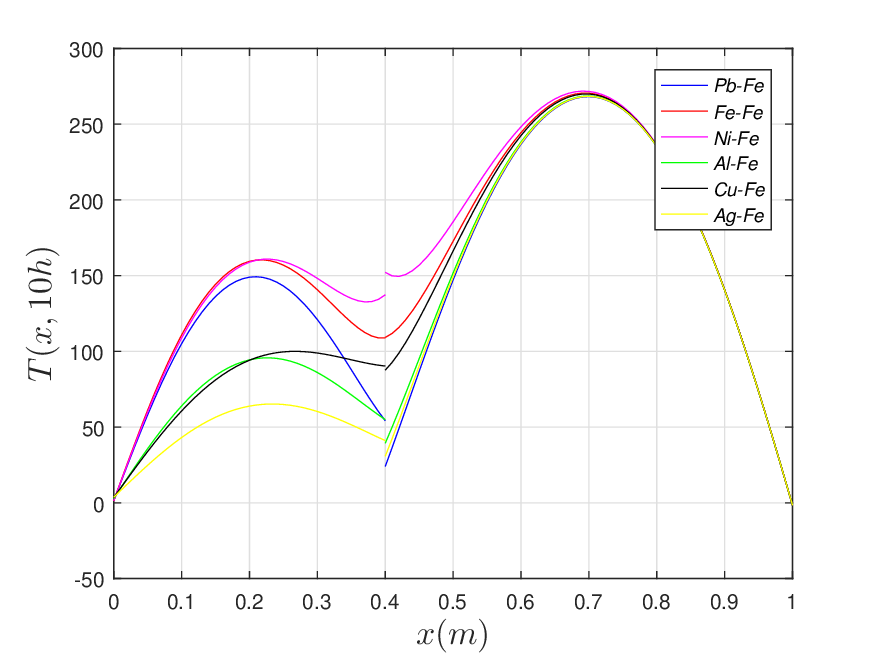}
\vspace{-0.5cm}
\caption{Temperature for $t=10h$, $Pb-Material$ (left) and $Material-Fe$ (right).}
\vspace{-0.5cm}
\label{Perfil_Esp_t_10}
\end{center}
\end{figure}
 
Figure \ref{Perfil_Esp_t_10} shows the spatial temperature profiles for $t=10 \, h$, where the temperature function
is defined by:
\begin{equation}
\label{upartida}
T(x,t)=
\begin{cases}
T_1(x,t), & \quad (x,t) \in [0,l] \times [0,t_\infty],  \\
T_2(x,t), & \quad (x,t) \in [l,L] \times [0,t_\infty].
\end{cases}
\end{equation}

On the left, the spatial temperature profiles for $Pb-Material$ are plotted, while on the right, the corresponding profiles for $Material-Fe$ are shown. In both situations, temperature discontinuities at the interface are observed; these discontinuities increase as the differences in thermal conductivity and diffusivity become larger. This observation aligns with the physics of the problem.
%
\begin{figure}[!h]
\begin{center}
\includegraphics[width=0.495\textwidth]{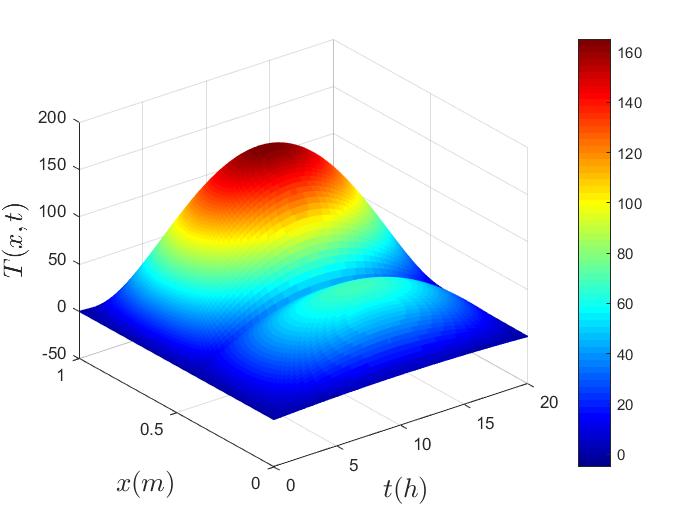}
\includegraphics[width=0.495\textwidth]{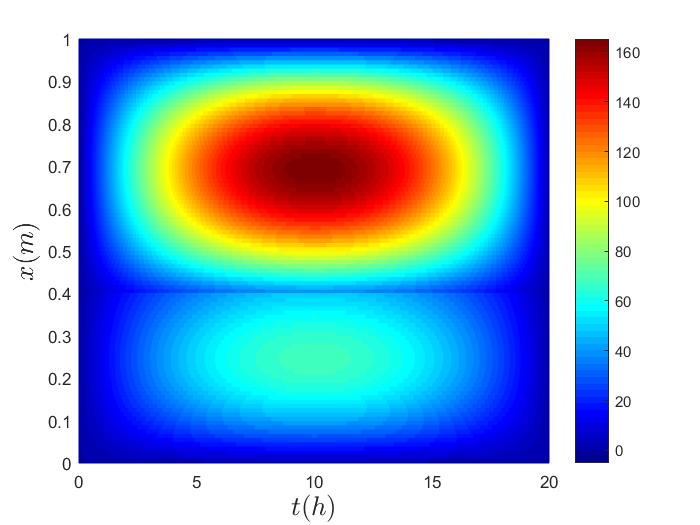}
\vspace{-0.5cm}
\caption{Temperature distribution for $Ag-Cu$.}
\vspace{-0.5cm}
\label{Esp_Tem_Ag_Cu}
\end{center}
\end{figure}

As an example, Figure \ref{Esp_Tem_Ag_Cu} shows the space-time temperature function for an $Ag-Cu$ body. It is observed that the maximum temperature slightly exceeds $160 \, ^{\circ}C$ and the temperature discontinuity is located at $x=0.4 \,m$, becoming more noticeable around $t=10 \, h$.

\begin{figure}[!h]
\begin{center}
\includegraphics[width=0.495\textwidth]{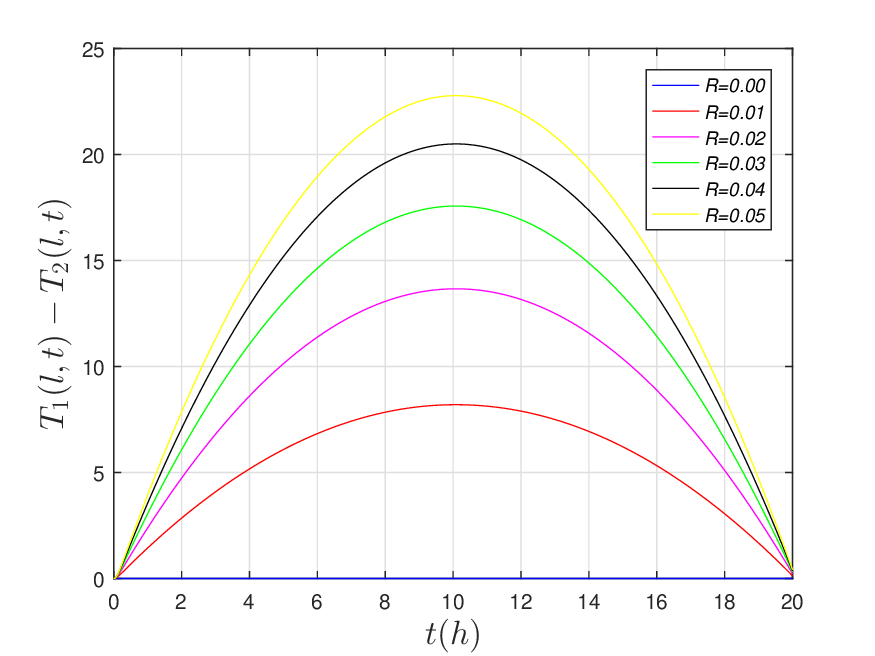}
\includegraphics[width=0.495\textwidth]{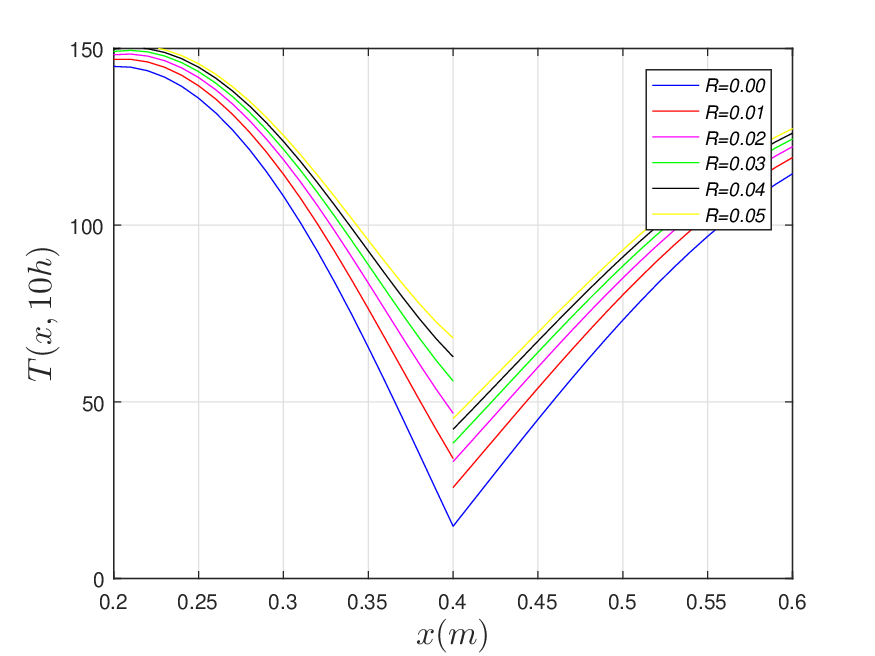}
\vspace{-0.5cm}
\caption{Distribution of temperature differences at the interface for $Pb-Ag$ considering different thermal resistances.}
\vspace{-0.5cm}
\label{dif_distint_R}
\end{center}
\end{figure}

Finally, Figure \ref{dif_distint_R} shows the spatial profile of the temperature difference at the interface for a $Pb-Ag$ body with different values of thermal resistance. It can be seen that the temperature difference increases as the resistance value rises. Additionally, the difference becomes more pronounced at $t=10 \, h$.
\begin{nt}

Since a stable and convergent numerical scheme is employed with an appropriate integration step, as demonstrated in the literature, similar configurations will indeed yield analogous results.

\end{nt}

\begin{nt}

The results presented in this article are applicable to any type of material, provided that the specified conditions and assumptions are upheld. This applicability is due to the fact that both the analytical and numerical solutions depend solely on the thermal conductivity and diffusivity coefficients of the materials.
\end{nt}

\section{Conclusions}

This paper presents a theoretical analysis of a one-dimensional heat transfer problem in a two-layer body. The analysis includes diffusion, advection, internal heat generation or loss that varies linearly with temperature in each layer, and heat generation from external sources. Additionally, the thermal resistance at the interface between the two materials is considered.
          
An analytical solution to the problem is derived using dimensionless variable transformations, differential equation techniques such as separation of variables, Fourier methods, and the principle of superposition. The analysis reveals that the associated eigenvalue equation has infinitely many solutions and the orthogonality condition is established. The analytical solution is shown to be consistent with previous literature for simpler cases, validating the approach used in this study.

Furthermore, a convergent finite difference method is proposed, incorporating a specialized treatment at the interface which results in a mixed finite difference scheme. This method effectively models the problem, offering valuable insights into temperature profiles and material behavior under varying conditions. 
The numerical results obtained align with the physical expectations of the problem. Specifically, the space-time temperature profiles exhibit a functional form similar to that of the source, and the behavior of different materials is consistent to their diffusivity and thermal conductivity: more diffusive materials exhibit a more rapid increase in temperature, while materials with higher thermal conductivity achieve higher temperatures.

\end{document}